\documentclass[prl,twocolumn,preprintnumbers,amsmath,amssymb,hyperref,letterpaper,superscriptaddress]{revtex4}
\usepackage{graphicx}
\usepackage{latexsym}
\usepackage{amsmath}
\usepackage{graphics}
\usepackage{amssymb}
\usepackage{layout}
\usepackage{verbatim}
\usepackage{amsfonts,epsfig}
\usepackage{slashed}
\usepackage{feynmp}
\usepackage{float}
\usepackage{bm}
\usepackage{bbm}
\usepackage{hyperref}
\usepackage{epstopdf}
\usepackage{xcolor}
\newcommand{\ket}[1]{\left|#1\right>}
\newcommand{\bra}[1]{\left< #1 \right|}
\newcommand{\beq}{\begin{equation}}
\newcommand{\eeq}{\end{equation}}
\newcommand{\bea}{\begin{eqnarray}}
\newcommand{\eea}{\end{eqnarray}}

\begin{document}

\unitlength = 1mm

\title{Fast microwave-driven three-qubit gates for cavity-coupled superconducting qubits}

\author{Edwin Barnes}
\affiliation{Department of Physics, Virginia Tech, Blacksburg, VA 24061, U.S.A.}
\author{Christian Arenz}
\affiliation{Department of Chemistry, Princeton University, NJ 08544}
\author{Alexander Pitchford}
\affiliation{Department of Physics, Aberystwyth University, UK}
\author{Sophia E. Economou}\email{economou@vt.edu}
\affiliation{Department of Physics, Virginia Tech, Blacksburg, VA 24061, U.S.A.}

\begin{abstract}
Although single and two-qubit gates are sufficient for universal quantum computation, single-shot three-qubit gates greatly simplify quantum error correction schemes and algorithms.
We design fast, high-fidelity three-qubit entangling gates based on microwave pulses for transmon qubits coupled through a superconducting resonator. We show that when interqubit frequency differences are comparable to single-qubit anharmonicities, errors occur primarily through a single unwanted transition. This feature enables the design of fast three-qubit gates based on simple analytical pulse shapes that are engineered to minimize such errors. We show that a three-qubit {\sc ccz} gate can be performed in 260 ns with fidelities exceeding $99.38\%$, or $99.99\%$ with numerical optimization.
\end{abstract}

\maketitle

Quantum information processing is one of the most exciting and rapidly growing fields of modern science, in large part due to quantum algorithms, which promise exponential speedup in solving important problems. Quantum two-level systems (qubits) are the fundamental carriers of quantum information, and among the most promising of these are qubits based on superconducting circuits \cite{Clarke_Nature08,Devoret_Science13}. Such an architecture is attractive because of the mature circuit fabrication technology and the ability to couple qubits together via resonators to implement logic gates \cite{Majer_Nature07}.

\begin{figure}[ptb]
\centering
\includegraphics[width=\columnwidth]{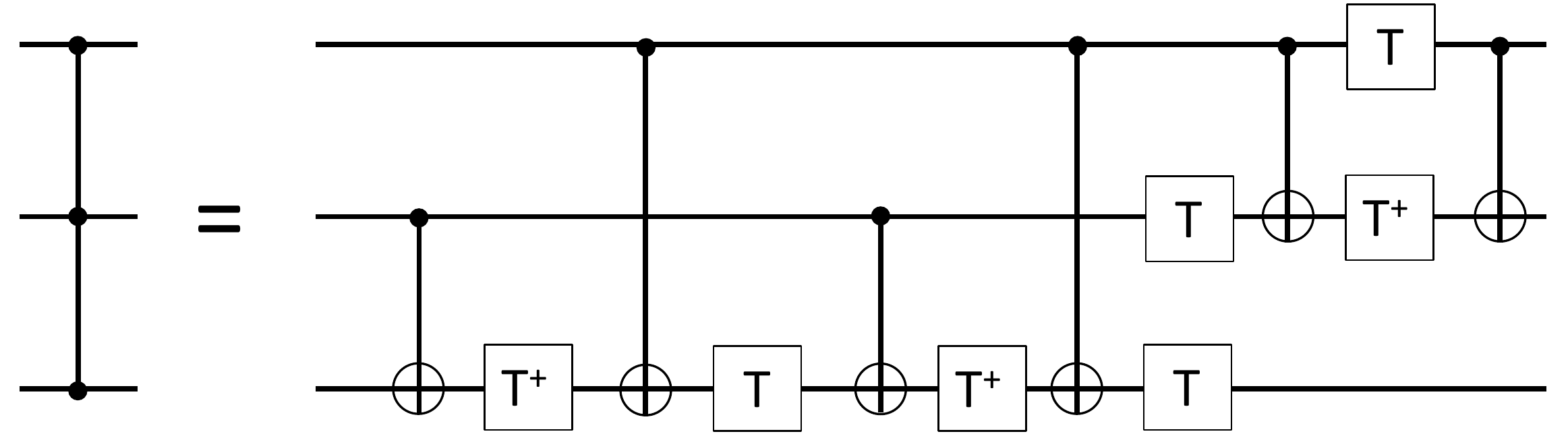}
\caption{Circuit diagram decomposition of the three-qubit {\sc ccz} gate in terms of seven single-qubit phase gates ($\top$) and six two-qubit {\sc cnot} gates.}
\label{fig:decomp}
\end{figure}

For universal quantum computing, a certain set of high-fidelity logic gates suffices to implement any algorithm; this set is comprised of single-qubit gates along with one maximally entangling two-qubit gate. Three-qubit gates, which play a prominent role in algorithms, can be decomposed in terms of a sequence of single- and two-qubit gates \cite{Nielsen_Chuang}. In the case of a maximally entangling three-qubit control-control-{\sc z} ({\sc ccz}) gate (described below), one must perform seven single-qubit gates and six entangling two-qubit gates, as shown in Fig.~\ref{fig:decomp}. The large number of gates needed makes it natural to explore whether a direct, single-shot three-qubit gate is preferable in terms of speed and fidelity.

There are two ways to implement logic gates in superconducting qubit systems. One is via tuning of the energy levels, which brings states into and out of resonance. The other is via oscillating microwave fields that induce transitions between energy levels, which offers the advantage of less susceptibility to charge noise. Both approaches face the challenge of spectral crowding in these systems, especially in the context of superconducting transmon qubits where each qubit is a weakly anharmonic oscillator out of which the two lowest levels are selected to encode information \cite{Koch_PRA07}. In the case of many qubits coupled together, there is a large number of closely spaced transitions that need to be avoided during quantum gate operations. Generically, a way to avoid unwanted transitions is to consider the time-energy uncertainty principle, which tells us to make the operations slow. In realistic systems, however, this is not an option, as we need to perform operations at a time scale that is much faster than decay and decoherence.

There has been substantial theoretical effort to design pulses that avoid unwanted dynamics due to `harmful' transitions. A protocol that avoids leakage in single-qubit gates called DRAG has been introduced \cite{Motzoi_PRL09,Forney_PRA10,Gambetta_PRA11,Motzoi_PRA13,Schutjens_PRA13} and experimentally used \cite{Chow_PRA10,Chen_PRL16}. For two-qubit gates, various approaches have been developed both for microwave drive \cite{Majer_Nature07,Li_PRB08,Kelly_PRL10,Rigetti_PRB10,Yang_PRA10,Kim_PRL11,Chow_PRL12,Chow_NJP13,Chow_NC14,Economou_PRB15} and for tuning \cite{Strauch_PRL03,DiCarlo_nature09,DiCarlo_nature10,Lucero_NP12,Ghosh_PRA13,Egger_SST14,Martinis_PRA14,Barends_Nature14}. In general, techniques developed in the context of one- or two-qubit gates are not directly applicable to three-qubit gates. Over the last few years researchers have used numerical optimal control theory (OCT) techniques to design three-qubit gates \cite{Stojanovic_PRB12,Zahedinejad_PRL15,Zahedinejad_PRApplied16,Moqadam_PRA13}. Thus far, only tuning-based three-qubit gates have been demonstrated experimentally \cite{Fedorov_Nature11,Mariantoni_science11,Reed_nature12}. Although many experimental groups are pursuing an all-microwave-control approach, microwave-based three-qubit gates have yet to be implemented in the laboratory, probably because there does not yet exist a simple protocol for them.

In this Letter, we present an analytical protocol for a fast, high fidelity three-qubit gate \cite{Economou_PRB15} based on a single microwave pulse. We show that by choosing qubit anharmonicities to be comparable to inter-qubit detunings, the spectrum resembles that of two qubits. In this regime, we can adapt a two-qubit gate protocol that two of us have recently developed \cite{Economou_PRB15}, known as SWIPHT, to implement a microwave-driven three-qubit gate in 260 ns. We use a hyperbolic secant pulse, which is smooth and experimentally simple to generate. Despite the complexity of the Hilbert space and the simplicity and speed of our pulse, we obtain a gate fidelity exceeding 99.38\%. We further show that our analytical solution can be successfully used as a starting point in OCT algorithms. The OCT solution reduces the infidelity by almost two orders of magnitude, pushing the gate fidelity to more than 99.99\% without affecting the total gate duration, but at the cost of a more complicated pulse.

The most well known three-qubit gate is the Toffoli gate, which is a {\sc not} operation on one qubit conditional on the state of the other two qubits, and which is also known as control-control-{\sc not} ({\sc ccnot}). The Toffoli gate is equivalent to a {\sc ccz} gate, which amounts to a {\sc z} gate (a $\pi$ rotation about the $z$ axis) on one of the qubits conditional on the state of the other two qubits. In the computational basis, the {\sc ccz} gate is a diagonal matrix with a single diagonal element equal to -1, and the rest equal to 1.

The Hamiltonian describing the three-qubit--cavity system is
\beq
H_0=\omega_{c}a^{\dagger}a+\sum_{\ell=1}^3[\omega_{\ell}b_{\ell}^{\dagger}b_{\ell}
+\frac{\alpha}{2}b_{\ell}^{\dagger}b_{\ell}(b_{\ell}^{\dagger}b_{\ell}-1)+g(a^{\dagger
}b_{\ell}+ab_{\ell}^{\dagger})].
\eeq
Here $a^{\dagger}(a)$, $b_\ell^{\dagger}(b_\ell)$ are creation
(annihilation) operators for the cavity and the transmons, respectively,
$\omega_\ell$ denotes the energy splittings between the first
excited state and ground state for each transmon, $\alpha$ is the anharmonicity, and $g$ is the coupling strength between each transmon and the cavity. We denote the non-interacting ($g=0$), `bare' eigenstates by $\ket{n;ijk}$, where $i,j,k=0,1,...$ label the individual transmon states, and where $n=0,1,...$ specifies the number of cavity photons. The eigenstates of the full, interacting system described by $H_0$ are denoted by $\widetilde{\ket{n;ijk}}$. When the qubits are detuned from each other and from the cavity---a typical situation for superconducting qubits---and for typical qubit-cavity coupling strengths on the order of tens or hundreds of MHz, each dressed eigenstate has a large overlap with one bare state. Accordingly, the energies of the eigenstates are shifted slightly from those of the bare states they originate from. Thus, we use bare state labels $n,i,j,k$ to label each interacting, dressed eigenstate according to the bare state with which it has the largest overlap. We define our computational states to be the lowest eight dressed states with zero cavity photons: $\widetilde{\ket{ijk}}$, with $i,j,k=0,1$. We omit the cavity photon label when it is zero.

\begin{figure}[ptb]
\centering
\includegraphics[width=0.7\columnwidth]{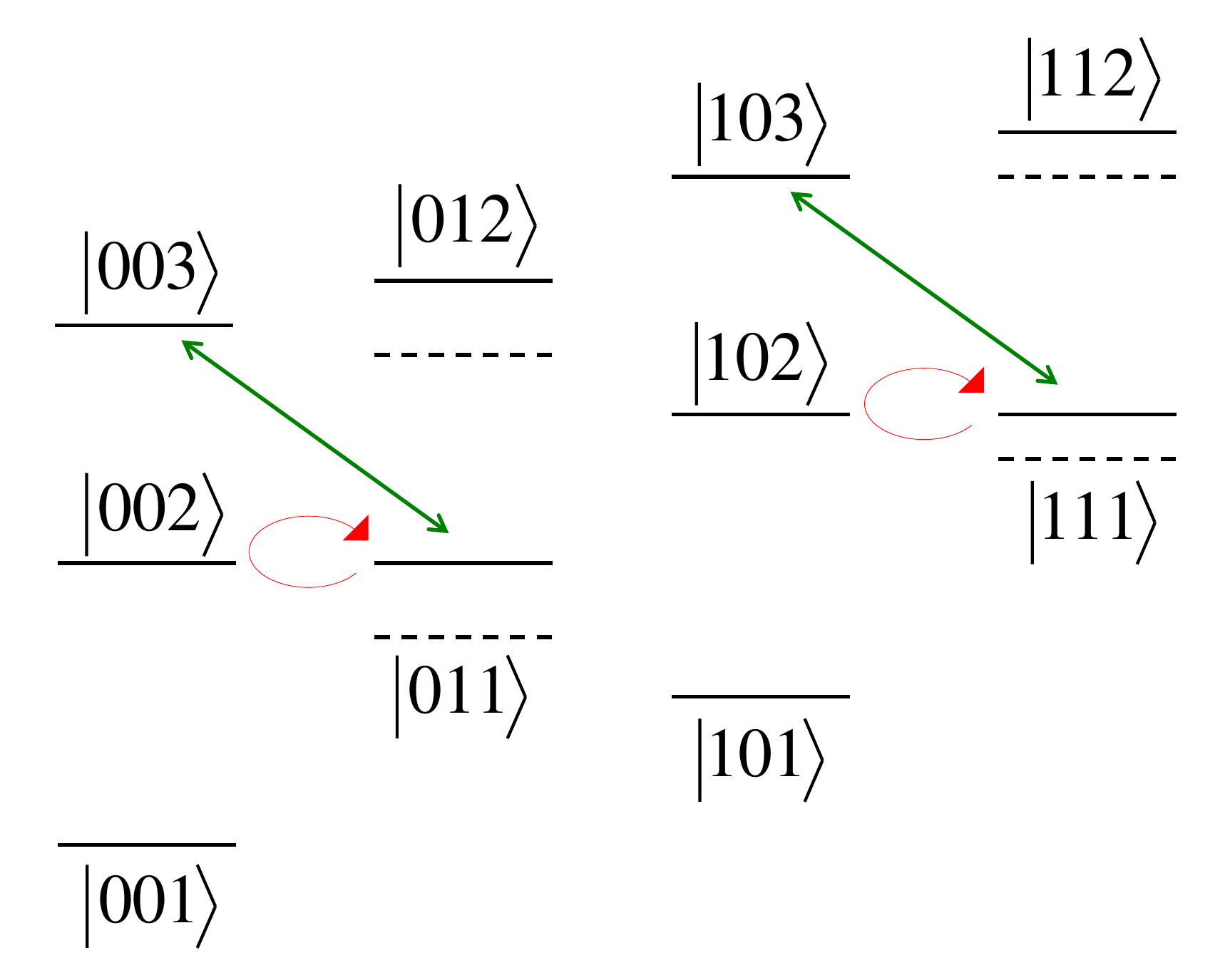}
\caption{(Color online) Subsector of the three-qubit bare state spectrum that is relevant when the $\ket{1}\leftrightarrow\ket{2}$ transition of transmon 3 is driven. For generic parameters (dashed levels), there are four nearly degenerate transitions corresponding to the four logical states of transmons 1 and 2. When the anharmonicity is close to the detuning between transmons 2 and 3 (solid levels), $\ket{011}$ and $\ket{002}$ become nearly degenerate and mix strongly, as do $\ket{111}$ and $\ket{102}$, enabling two nearly degenerate two-qubit-like transitions, $\widetilde{\ket{011}}\leftrightarrow\widetilde{\ket{003}}$ and $\widetilde{\ket{111}}\leftrightarrow\widetilde{\ket{103}}$ (green arrows). Dressed state tildes have been suppressed for clarity.}
\label{fig:spectrum}
\end{figure}

Throughout this work, we focus on the case in which only one transmon is driven, and we take this to be transmon 3. In this case, the full Hamiltonian with driving included is
\beq
H(t)=H_0+b_3\Omega(t)e^{i\omega_p t}+b_3^\dagger\Omega(t)e^{-i\omega_pt},
\eeq
where $\Omega(t)$ and $\omega_p$ are the pulse envelope and frequency, respectively. The diagonal nature of the {\sc ccz} gate immediately points to transitionless evolution, i.e., evolution generated by pulses that transfer population between energy eigenstates only transiently and induce phases on the eigenstates. There is a fundamental challenge in inducing a minus sign to only one of the eigenstates while not affecting the remaining ones, and it is related to the generic structure of a three-qubit system.  Specifically, it is clear that when we drive one of the three qubits, e.g., transition $|1\rangle \leftrightarrow |2\rangle$ of qubit 3, there are four states that are near-degenerate:  $\widetilde{|ij1\rangle} \leftrightarrow \widetilde{|ij2\rangle}$, with $i,j$ describing the 0,1 states of the other two qubits (see Fig.~\ref{fig:spectrum}). If we attempt to implement the {\sc ccz} gate by simply selecting narrow-band pulses to resolve the target transition without disturbing the three remaining near-degenerate  transitions, we will end up with extremely long pulses.

To speed up the pulse, we employ the basic concept of the SWIPHT method, which we introduced in \cite{Economou_PRB15}. The main idea behind SWIPHT is to stop trying to avoid the nearly-degenerate transitions, but to instead purposely drive them using a pulse that is properly designed so that the net effect is that the driven states acquire trivial phases. This idea was utilized in Ref. \cite{Economou_PRB15} in the context of two cavity-coupled qubits, for which there are two near-degenerate transitions, to implement both the {\sc cnot} and the {\sc cz} gate. In the case of the {\sc cz} gate, which is the two-qubit analogue of our three-qubit {\sc ccz} gate, a pulse of hyperbolic secant pulse shape was used, which yields an analytically solvable Schr\"{o}dinger equation \cite{Rosen_PR32} for a two-level system. The fact that in the two-qubit case there are two transitions driven simultaneously in a prescribed way allowed us to tune two parameters, the detuning and bandwidth of the pulse, to induce the correct phases on each eigenstate \cite{Economou_PRB06,Economou_PRL07,Economou_PRB12}.

In the present setup, the protocol of Ref. \cite{Economou_PRB15} cannot be used straightforwardly due to the existence of the four $\widetilde{|ij1\rangle} \leftrightarrow \widetilde{|ij2\rangle}$ transitions: the number of pulse parameters is simply less than the number of constraints. We address this issue by taking a second look at the spectrum to explore whether a more advantageous structure can be achieved by appropriately choosing the static parameters, i.e., the qubit and cavity frequencies, the anharmonicity, and the coupling strength (we assume the latter two are the same for all qubits). Out of the four transitions $\widetilde{|ij1\rangle} \leftrightarrow \widetilde{|ij2\rangle}$, we aim to isolate two as near-degenerate so that the spectrum will resemble the two-qubit spectrum, which we know how to control. If we choose the frequency of the qubits to differ by the anharmonicity, then the bare state $|011\rangle$ is degenerate with $|002\rangle$ and $|111\rangle$ is degenerate with $|102\rangle$, which means that the two states $|011\rangle$ and $|111\rangle$ will mix more strongly with the excited states $|002\rangle$ and $|102\rangle$. As a result, when the third transmon qubit is driven, there will be a significant dipole matrix element between the states $\widetilde{|011\rangle}$ and $\widetilde{|003\rangle}$ and between the states $\widetilde{|111\rangle}$ and $\widetilde{|103\rangle}$. Transitions near these frequencies involving the remaining two states, $\widetilde{|001\rangle}$ and $\widetilde{|101\rangle}$, will not be present (see Fig.~\ref{fig:spectrum}).

\begin{figure}[!h]
\vspace{-0.5cm}
\includegraphics[width=\columnwidth]{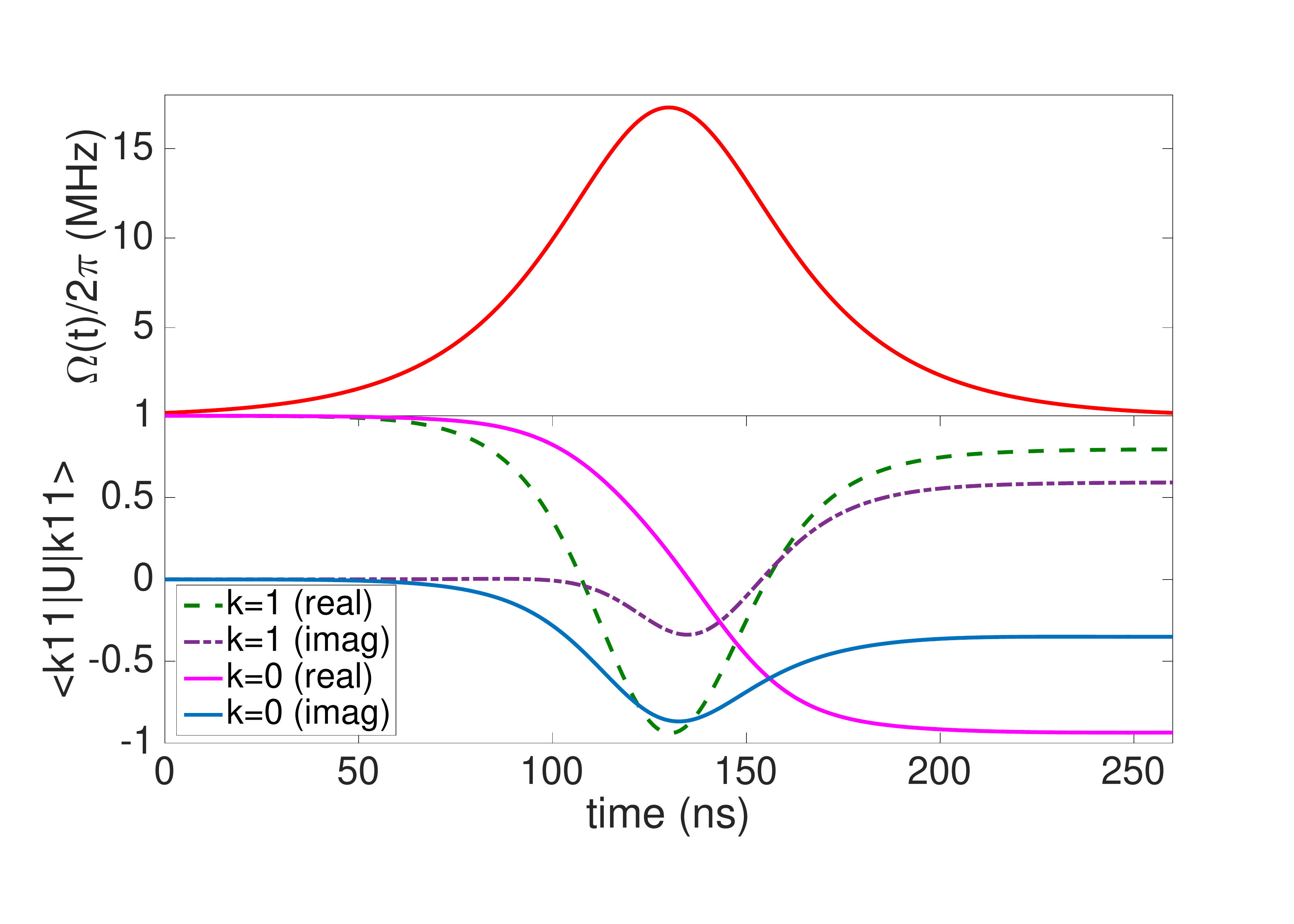}\\
\vspace{-0.5cm}
\caption{(Color online) Upper panel: The hyperbolic secant pulse, $\Omega(t)=\Omega_0\hbox{sech}(\sigma(t-T/2))$, used to perform the {\sc ccz} gate. Here, $\sigma=\delta/\sqrt{3}=6.12$ MHz, $T=260.02$ ns, $\Omega_0=2\sigma/d=17.3$ MHz, and $d=0.7074$ is the ratio of the dipole moments of the non-interacting single-qubit $\ket{0}\leftrightarrow\ket{1}$ transition and interacting harmful transition. This choice of $\Omega_0$ implements a $4\pi$ rotation on the harmful transition. The simulations were performed using the following parameters: $\omega_1=6.2$ GHz, $\omega_2=6.6$ GHz, $\omega_3=7.0$ GHz, $\omega_c=7.15$ GHz, $\alpha=350$ MHz, $g=130$ MHz. Lower panel: Simulation of the evolution operator matrix elements corresponding to the logical states involved in the target ($k=0$) and harmful ($k=1$) transitions. The target state $\widetilde{\ket{011}}$ picks up a $\pi$ phase, while the state $\widetilde{\ket{111}}$ picks up a trivial $2\pi$ phase. Deviations from the ideal behavior are largely due to correctable local operations.}\label{fig:evolop}
\end{figure}

With this spectrum, we are now able to adapt the SWIPHT protocol. We choose the two transitions, $\widetilde{|011\rangle}\leftrightarrow \widetilde{|003\rangle}$ and
$\widetilde{|111\rangle}\leftrightarrow \widetilde{|103\rangle}$, as the target and the harmful transitions, respectively. We select a hyperbolic secant pulse of area 4$\pi$, which acts on both transitions simultaneously (see Fig.~\ref{fig:evolop}). Applying this pulse resonantly to the harmful transition will drive it strongly, but will only induce a trivial phase of $2\pi$ on it. This result is independent of the pulse bandwidth, $\sigma$, which is a free parameter that we can adjust to ensure that the phase acquired by the target transition is -1 as needed for the {\sc ccz} gate. The analyticity of the solution leads to an explicit result for this phase: $\phi=2\arctan(4\delta\sigma/[\delta^2-3\sigma^2])$, where $\delta$ is the detuning between the two transitions \cite{Economou_PRB12}. The value of the bandwidth that achieves $\phi=\pi$ is $\sigma=\delta/\sqrt{3}$.

So far, our discussion ignores the rest of the Hilbert space and assumes that the presence of the large number of states and transitions not directly involved in our protocol will not affect our result. We now proceed to test this by simulating the dynamics of the full system to assess the performance of our solution in a realistic setup. The system and pulse parameters we use are shown in the caption of Fig.~\ref{fig:evolop}. The pulse envelope is shown in the upper panel of the figure, where it can be seen that the total duration is $T=260$ ns. In the simulations, we retain enough states out of the infinite Hilbert space to ensure that the results converge, which amounts to keeping 4 cavity states and 5 states for each of the qubits, resulting in a 500-dimensional Hilbert space. The matrix elements of the evolution operator corresponding to the logical states involved in the target and harmful transitions are shown in the lower panel of Fig.~\ref{fig:evolop}. Although deviations from the ideal values $\bra{011}U(T)\ket{011}=-1$ and $\bra{111}U(T)\ket{111}=1$ are evident, we will see that these deviations are largely due to the pulse generating additional local operations that can be corrected with single-qubit gates.

\begin{figure}[!h]
\includegraphics[width=\columnwidth]{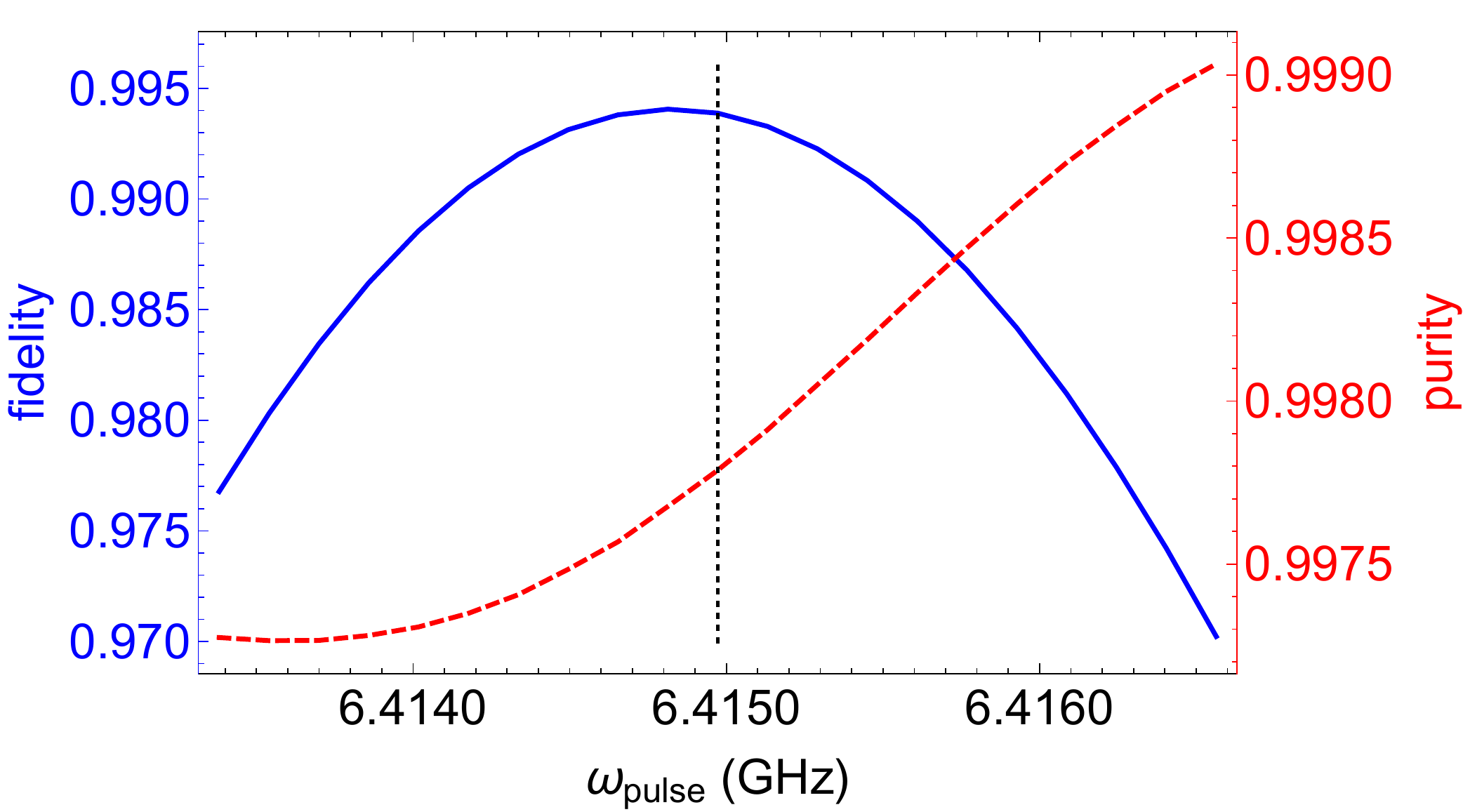}\\
\caption{(Color online) Fidelity (solid blue curve, left scale) and purity (dashed red curve, right scale) as a function of pulse frequency. The vertical dotted line indicates the frequency of the harmful transition.}\label{fig:fidelity}
\end{figure}

We quantify the performance of our {\sc ccz} gate using the fidelity defined in Ref.~\cite{Pedersen_PLA07},
\begin{equation}
F=\frac{1}{72}[Tr(MM^{\dagger})+|Tr(M)|^{2}],
\end{equation}
where $M=U_{ideal}U^{\dagger}$, $U$ is the actual evolution operator truncated to the three-qubit logical subspace, and $U_{ideal}$ is the target gate operation. To account for possible local operations, we choose $U_{ideal}$ to be a generalized {\sc ccz} gate which is locally equivalent to the standard {\sc ccz} up to single-qubit phase gates and a global phase: $U_{ideal}=e^{-i\phi_0-i\sum_\ell\phi_\ell\sigma_{\ell,z}}\hbox{diag}(1,1,1,-1,1,1,1,1)$, and we maximize $F$ over the phases $\phi_k$. We find that our gate has a fidelity of $F=0.993881$ with phases $\phi_0=1.19187$, $\phi_1=0.03607$, $\phi_2=-0.02842$, $\phi_3=0.19056$. Considering the complexity and spectral crowding of the Hilbert space and the simplicity of our pulse shape, this is a remarkable, and perhaps somewhat surprising result. For comparison, even if it were possible to perform every single- and two-qubit gate in the {\sc ccz} decomposition shown in Fig.~\ref{fig:decomp} with 99.9\% fidelity, the fidelity of the resulting {\sc ccz} gate would be $\sim98.7$\%, while the total duration would exceed a microsecond. The fidelity and purity as a function of the pulse frequency are shown in Fig.~\ref{fig:fidelity}, where it is evident that the fidelity can be pushed a little higher by shifting the frequency slightly away from resonance with the harmful transition. It is likely that still higher fidelities can be achieved by optimizing with respect to other pulse or system parameters. It is also apparent that the purity grows steadily with pulse frequency, indicating that a larger portion of the error is contained within the logical subspace.

We now address the issue of further increasing the gate fidelity via OCT by using our analytical pulse as a starting point. The idea of OCT is to maximize or minimize a given cost functional based on the Pontryagin maximum principle. Typically this is done by numerically searching for global optima using a gradient based approach such as gradient ascent pulse engineering (GRAPE) \cite{Khaneja_JMR05}, which is implemented in the control package in QuTiP \cite{qutip2,qutip4}. Here we use the GRAPE algorithm in order to find the pulse $\Omega(t)$ that maximizes the fidelity. If we divide the total evolution time $T$ into equispaced time intervals, $\Delta t=T/n$, on which $H(t)$ is assumed to be piecewise constant, the total evolution $U(T)$ can be written as a time order product of unitary operators:
\begin{align}
	U(T)=\prod_{j=1}^{n}e^{-iH(t_{j})\Delta t},
\end{align}
 where $t_{j}\in[(j-1)\Delta t,j\Delta t]$. Denoting by $\Omega_{1},\ldots,\Omega_{n}$ the piecewise constant pulse amplitudes, it is straightforward to calculate the gradient of $F$ with respect to $\Omega_{i}$,  noting that an expression for $\frac{\partial U(T)}{\partial \Omega_{i}}$ can be found in \cite{Nigmatullin_NJP09}. After having modified the gradient expression in \cite{qutip2,qutip4} accordingly, we maximized $F$ for a total evolution time $T=260.06\,\text{ns}$ using $n=500$ time slots and taking the generalized {\sc ccz} gate with pre-optimized phases $\phi_k$ as the target gate.

 \begin{figure}
\includegraphics[width=0.9\columnwidth]{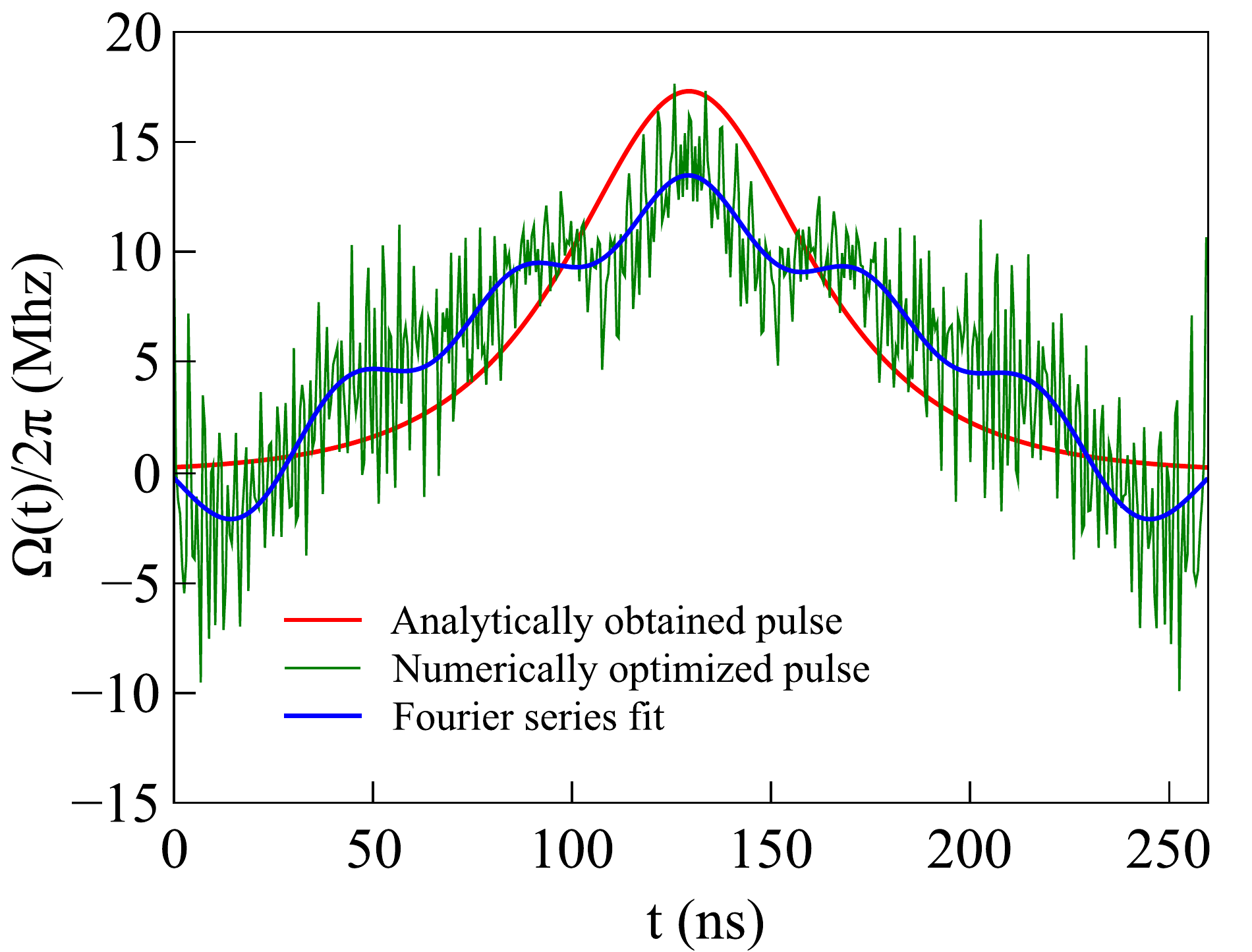}\\
\caption{(Color online) Pulse envelope implementing the {\sc ccz} gate as a function of time. The smooth red curve shows the analytically obtained pulse corresponding to a fidelity of $F=0.993881$. The rapidly modulated green curve shows the numerically optimized pulse using OCT, which further decreases the fidelity by two orders of magnitude ($F=0.999941$). The more slowly modulated blue curve shows a Fourier series fit of the numerically optimized pulse with fidelity $F=0.995574$.}\label{fig:oct}
\end{figure}

The results are shown in Fig.~\ref{fig:oct}. As an initial guess pulse we took the $4\pi$ hyperbolic secant pulse with $\sigma=\delta/\sqrt{3}$ (smooth red curve), which has the advantage of being already close to the optimal solution. The rapidly modulated (green) curve shows the optimized pulse corresponding to a fidelity of $F=0.999941$ and the slowly modulated (blue) curve shows a Fourier series fit with seven coefficients yielding $F=0.995574$. Thus, starting from the analytically obtained pulse, the use of OCT can further decrease the infidelity by a few orders of magnitude. We find that using simple, near-optimal analytical pulses as a seed in the numerical optimization algorithm can be dramatically more efficient than blind numerical searches. On the other hand, Fig.~\ref{fig:oct} suggests that pulses yielding higher fidelities contain significantly higher frequency components, and attempts to smoothen out the fast oscillations lead to a quick saturation of the fidelity to the value obtained for the initial analytical pulse.

In conclusion, we have designed a microwave-based single-shot three-qubit gate for transmon qubits generated by a simple, fast and smooth pulse. Our gate is based on selecting system parameters to obtain a more favorable spectrum, reminiscent of a two-qubit--cavity system. Even without optimizing over parameters or pulse shape, the gate fidelity comfortably exceeds 99\%. We have demonstrated that using our analytical pulse as a starting point in optimal control theory leads to a further decrease in the gate infidelity by two orders of magnitude.

\end{document}